# Chaos and noise in a truncated Toda potential

Salman Habib*, Henry E. Kandrup†‡, and M. Elaine Mahon†

*T-6, Theoretical Astrophysics
and
T-8, Elementary Particles and Field Theory
Los Alamos National Laboratory
Los Alamos, New Mexico 87545

†Department of Astronomy and Institute for Fundamental Theory
University of Florida
Gainesville, Florida 32611

## Abstract

Results are reported from a numerical investigation of orbits in a truncated Toda potential which is perturbed by weak friction and noise. Two significant conclusions are shown to emerge: (1) Despite other nontrivial behaviour, configuration, velocity, and energy space moments associated with these perturbations exhibit a simple scaling in the amplitude of the friction and noise. (2) Even very weak friction and noise can induce an extrinsic diffusion through cantori on a time scale much shorter than that associated with intrinsic diffusion in the unperturbed system.

e-mail:
habib@eagle.lanl.gov
kandrup@astro.ufl.edu
mahon@astro.ufl.edu
‡Also *Department of Physics, University of Florida*

In the past several years, substantial interest has centered on the study of stochastic orbits in nonintegrable Hamiltonian systems. In particular, much work has focused on the phenomenon of intrinsic diffusion through cantori, *e.g.*, using the "turnstile" model of MacKay *et al* [1], which leads to interesting scaling behaviour [2], and, more recently [3], through the utilisation of local Lyapunov exponents [4]. However, with a few notable exceptions (cf. [5]), little has been done to determine how the results derived from such analyses are changed if non-Hamiltonian perturbations are allowed. This is an issue of key importance given the modern perspective on the connection between dynamical chaos and the foundations of classical statistical mechanics [6].

In modeling a physical system in terms of a few degree of freedom Hamiltonian, one is usually implementing a reduced description, *e.g.*, in terms of collective coordinates, that neglects (hopefully) weak couplings to an external environment and/or various (relatively) unimportant degrees of freedom. However, such weak corrections can oftentimes be viewed as a source of friction and noise, related via a fluctuation-dissipation theorem, and, as such, it is important to investigate the qualitative and quantitative effects that arise if the Hamiltonian model be perturbed to allow for weak friction and noise. Naively, one might expect that weak coupling to an environment should only have effects on very long time scales. However, as will be shown here, for nonintegrable systems such effects can in fact be important on time scales much shorter than the natural relaxation time scale.

The system considered here in detail is the sixth order truncation of the Toda [7] potential, which leads to the Hamiltonian

$$H = \frac{1}{2}\left(v_x^2 + v_y^2\right) + \frac{1}{2}\left(x^2 + y^2\right) + x^2y - \frac{1}{3}y^3 + \frac{1}{2}x^4 + x^2y^2 + \frac{1}{2}y^4$$

$$+ x^4y + \frac{2}{3}x^2y^3 - \frac{1}{3}y^5 + \frac{1}{5}x^6 + x^4y^2 + \frac{1}{3}x^2y^4 + \frac{11}{45}y^6, \quad (1)$$

where $\{x, v_x\}$ and $\{y, v_y\}$ represent conjugate pairs. The Hamiltonian evolution was perturbed by allowing for a constant friction $-\eta\mathbf{v}$ and delta-correlated additive white noise $\mathbf{F}(t)$, related via a fluctuation-dissipation theorem with temperature $\Theta \sim E$. Other forms of friction and noise may give different conclusions, a possibility currently under investigation.

For each initial condition, an unperturbed Hamiltonian trajectory was computed. Multiple Langevin simulations were then performed, using a numerical algorithm (cf. [8]) which generates a random $\mathbf{F}$ with the proper first and second moments. A time step $h = 0.001$ was used for most simulations and it was verified that smaller time steps led to no statistically significant differences. The total time for each integration was $t \leq 140$ (the dynamical or crossing time is of order unity).

Viewed over long time scales, the unperturbed Hamiltonian trajectories divide naturally into only two classes, namely regular orbits, with vanishing Lyapunov exponent, and stochastic orbits, with nonvanishing Lyapunov exponent. However, on shorter time scales, the stochastic orbits divide in turn into two relatively distinct types, namely filling stochastic orbits which travel unimpeded throughout the stochastic regions and confined, or sticky, stochastic orbits, which are trapped near islands of regularity by cantori, and only escape over much longer time scales. It is therefore meaningful to consider the effects of friction and noise separately on these three different orbit classes.

Experiments focused on the energy range $10 \leq E \leq 100$, with $10^{-12} \leq \eta \leq 10^{-3}$ and $0.1 \leq \Theta/E \leq 10.0$. Two different types of experiments were performed:

(1) For each of a large number of individual initial conditions, corresponding to both regular and stochastic orbits, $N = 48$ different noisy realisations were effected. The orbits were compared to the unperturbed trajectories and analysed statistically to extract the first and second moments of such quantities as the position, velocity, and energy, e.g.,

$$\langle |\delta E|^2 \rangle \equiv \delta E_{rms}^2 \equiv \frac{1}{N} \sum_{i=1}^{N} (E_{unp} - E_i)^2. \qquad (2)$$

(2) The near-invariant distribution associated with the Hamiltonian evolution was sampled to extract 400 initial conditions corresponding to stochastic orbits, and 100 noisy realisations were effected for each of these initial conditions. The outputs were then analysed in two ways, (a) by comparing with the unperturbed trajectories to extract first and second moments which average over the ensemble of initial conditions, and (b) by binning the orbital data at fixed time intervals to study systematic changes in the form of the near-invariant distribution. This latter set of experiments was performed for $E = 30$ and $75$ with $\Theta = E$ and $\eta = 10^{-9}$, $10^{-6}$, and $10^{-4}$. A single experiment was done with $E = \Theta = 10$ and $\eta = 10^{-6}$. Each of these latter experiments took approximately 250 node hours on the Los Alamos CM5.

Viewed in energy space, weak friction and noise serve to induce a classical diffusion process. Specifically, at least for early times, when $\delta E_{rms}/E_{unp} \ll 1$, $\delta E_{rms}$ satisfies the simple scaling relation

$$\delta E_{rms}^2 = A^2(E) E \eta \Theta t, \qquad (3)$$

where $A(E)$ is only weakly dependent on $E$. This scaling holds for all three classes of orbits, regular, sticky stochastic, and filling stochastic. Moreover, it holds for individual initial conditions as well as for ensembles of initial conditions. Viewed in energy space, one cannot distinguish between different orbit classes.

While for individual initial conditions, $A$ exhibits some variability ($\sim 10\%$), when averaged over the ensembles of initial conditions, the best fit value is very well determined. For $E = 75$, $1.68 < A < 1.71$, for $E = 30$, $1.66 < A < 1.69$, and, for $E = 10$, $1.66 < A < 1.67$. There is no systematic residual dependence on $\eta$. The results suggest that, when $E$ is decreased, the best fit value of $A$ also decreases, albeit very slowly. In the limit $E \to 0$, orbits essentially move in a harmonic oscillator potential, and one would anticipate an asymptotic approach to the exact oscillator result, $\delta E_{rms}^2 = 2E\eta\Theta t$, i.e., $A = \sqrt{2}$. (This result holds in the limit $\eta \ll 1$, assuming "orbit averaging," so that, e.g., $\sin^2 t \to 1/2$.) The data are consistent with this expectation.

Viewed in configuration or velocity space, friction and noise have more complicated effects, the form of which depend on the class of orbit, i.e., whether regular, sticky stochastic, or filling stochastic. For regular orbits the second moments in position and velocity grow as a power law in $t$, albeit more rapidly than the analytically predicted relation $\delta x_{rms}, \delta v_{x,rms}, ... \propto t^{1/2}$ which is satisfied by the integrable cases of a harmonic oscillator or a free particle. By contrast, for stochastic orbits these *rms* quantities grow exponentially at a rate $\Lambda$ that is comparable in magnitude to the Lyapunov exponent $\chi$, which characterises the average instability of the Hamiltonian orbit (See Fig. 1). This conclusion holds both for entire ensembles and for individual initial conditions. In the latter case, one also observes a direct correlation between the growth rate $\lambda_i$ for an individual initial condition and the local Lyapunov exponent $\chi_i$ for that initial condition. This correlation is particularly strong for the case of weak friction and noise. Moments for the confined stochastic orbits exhibit an intermediate behavior.

Despite these differences, one observes a universal characteristic relating the moments for all three classes of orbits, namely a simple scaling in terms of $\Theta$ and $\eta$. Specifically, provided that the deviations have not become macroscopic (i.e., assuming $\delta x_{rms}, \delta y_{rms} \ll 1$), all three types of orbits satisfy a scaling relation

$$\delta x_{rms}, \delta y_{rms}, \delta v_{x,rms}, \delta v_{y,rms} \propto \Theta^a \eta^b F(E,t), \qquad (4)$$

where $a = b = 0.50 \pm 0.01$ for the range of values probed. This was confirmed both for multiple realisations of individual initial conditions and for ensembles of initial conditions corresponding to stochastic orbits. It follows that, even for stochastic orbits, one observes the same scaling in $\Theta$ and $\eta$ as for regular orbits in a harmonic oscillatory potential, although the time dependence is extremely different.

Another common feature of all three orbit types is that, at sufficiently early times, the time dependence is reasonably well fit by a power law, i.e.,

$$\delta x_{rms}, \delta y_{rms}, \delta v_{x,rms}, \delta v_{y,rms} \propto \tilde{F}(E)\Theta^a \eta^b t^c. \qquad (5)$$

The interval over which such a fit is appropriate depends on the orbit type: filling stochastic orbits rapidly begin to show an exponential divergence, confined stochastic orbits only somewhat later, and regular orbits continue to manifest an approximate power law growth. The best fit value of $c$ depends on the sampling interval: for all types of orbits, if one fits to longer time intervals (again before the perturbation has become macroscopic), the best fit value of $c$ increases, even for the case of regular orbits. Fitting for relatively short times yields a value $c \approx 1.10 - 1.25$. Fitting regular orbits over longer times yields $c \approx 1.20 - 1.50$.

To test whether the early time scaling might be a numerical artifact, e.g., due to the finite step size, simulations were performed for a harmonic oscillator potential, selecting $E = \Theta = 30$ and $\eta = 10^{-6}$. The second moments in position and velocity were found to be in complete agreement with the analytic predictions for the moments $\delta x_{rms}, \delta y_{rms}, \delta v_{x,rms}$, and $\delta v_{y,rms}$.

To investigate the generality of these scaling relations, simulations were also performed for the dihedral $D4$ potential of Ref. [9], with a choice of parameter values which leads to a large amount of stochasticity. Specifically, the Hamiltonian was taken to be of the form

$$H = \frac{1}{2}(v_x^2 + v_y^2) - \frac{M}{2}(x^2 + y^2) - \frac{a}{4}(x^2 + y^2)^2 - \frac{b}{2}x^2 y^2, \qquad (6)$$

with $M = 2$, $a = -1$, and $b = 1/2$.

Analysis of these simulations shows once again a scaling of the form given by (3) and (4), with the same values $a = b = 0.50 \pm 0.01$. The best fit $c$ for Eqn. (4) shows more variability from run to run, so that a detailed comparison is difficult to effect. However, the data are consistent with the conclusion that $c$ is the same for both models.

When viewed in energy space, friction and noise again induce a simple diffusion process of the form (1) for the $D4$ potential. However, in this case the value of $A(E)$ shows somewhat more variability. For $1 \leq E \leq 50$, $A$ varies from $\sim 1.6 - 2.6$, with the largest values correlating with energies where, as characterised by the magnitude of the Lyapunov exponent, the phase space is most chaotic.

Generic ensembles of initial conditions corresponding to filling stochastic orbits, when evolved with the truncated Toda Hamiltonian (1), exhibit a coarse-grained exponential evolution towards an approximately invariant distribution [3]. For high energies $E \geq 50$, where the regular regions are very small, this distribution appears to correspond to a true invariant measure. However, for lower energies, where the relative measure of the regular regions is larger, this distribution slowly changes in form over longer time scales. As a result of intrinsic diffusion, the orbits can pass through the cantori to occupy phase space regions near the regular islands from which they were excluded at earlier times. However, friction and noise can significantly accelerate the changes in this near-invariant distribution by serving as a source of extrinsic diffusion. This has important implications for problems in galactic dynamics, which will be discussed elsewhere [10].

The Hamiltonian evolution conserves energy, restricting orbits to a constant energy hypersurface. When friction and noise are included, the energy is no longer conserved. However, it is approximately conserved over sufficiently short time scales, so that one can still speak of an "almost constant energy hypersurface." It is therefore meaningful to quantify the degree to which friction and noise alter the form of the near-invariant distribution on a near-constant energy hypersurface.

At a coarse-grained level, an ensemble of orbits can be characterised numerically by binned, projected distributions, such as $f(x, y, t; \Delta t)$ or $f(y, v_y, t; \Delta t)$, constructed by averaging the binned orbital data over some time interval $\Delta t$ [3]. In order to compare two different distributions $f_1$ and $f_2$, one requires a notion of distance. This was provided through the introduction of a coarse-grained $L^1$ norm: For two identically normalised distributions, $f_1(x,y)$ and $f_2(x,y)$, binned in an $n \times n$ grid of cells of size $\{\Delta x, \Delta y\}$,

$$Df_{1,2}(x,y) = \frac{\sum_{i=1}^{n}\sum_{j=1}^{n} |f_1(x,y) - f_2(x,y)|}{\sum_{i=1}^{n} f_1(x,y)}. \qquad (7)$$

It is with respect to this measure of distance that the Hamiltonian flow evidences an evolution towards a near-invariant measure.

However, if this near-invariant measure be evolved into the future, allowing for even weak friction and noise, one can observe systematic changes on time scales much shorter than the intrinsic diffusion timescale. The friction and noise can induce a significant extrinsic diffusion which allows filling stochastic orbits to become confined, thereby populating regions of the phase space near regular islands which were avoided by the deterministic near-invariant measure.

For the Toda potential, the observed changes are larger for lower energies, where the regular regions comprise a bigger fraction of the available phase space, and they occur more rapidly when the friction and noise are larger in amplitude. The case of $E = 30$, where cantori are very important, is particularly illuminating. For $\eta = 10^{-6}$, there is clear evidence that the noisy ensemble is evolving towards a new distribution which, over time scales $t \sim 100$, is approximately time-independent. For $\eta = 10^{-9}$, the ensemble again evidences a non-trivial time evolution, but the changes are sufficiently slow that one does not see clear evidence of an approach towards a new time-independent form. For $\eta = 10^{-4}$, changes in the energy are so large that it no longer makes sense to speak of an approximately constant energy hypersurface. However, the data are still consistent with an evolution towards a modified distribution in which cantori around the regular regions have been breached.

This behaviour is illustrated for $E = 30$ in Fig. 2. The first panel shows the form of the near-invariant distribution $f_0(x, y)$ associated with the deterministic evolution. The second panel shows the form of the noisy near-invariant distribution $f_\eta(x, y)$ for $\eta = 10^{-6}$. The final panel exhibits the difference $f_0(x, y) - f_\eta(x, y)$. The deterministic distribution $f_0(x, y)$ has four sharp relative minima which have become blurred somewhat in the noisy $f_\eta(x, y)$ by the diffusion of orbits into lower density regions.

At least for $E = 30$, the evolution from $f_0$ to $f_\eta$ is well fit by an exponential, with a rate $\Lambda$ that scales roughly as $\log \eta$. For a $10 \times 10$ binning, the best fit values for the slope are: for $\eta = 10^{-9}$, $\Lambda = -0.0143$; for $\eta = 10^{-6}$, $\Lambda = -0.0339$; and for $\eta = 10^{-4}$, $\Lambda = -0.0460$.

The fact that orbits can pass through cantori, going either in or out, implies the possibility of changes in orbit class between filling and confined stochastic orbits. It is difficult to construct a simple numerical algorithm to decide when an orbit has changed class. However, visual inspection of a large number of orbits ($\sim 10^4$) leads to qualitative conclusions that are easily summarised. At high energies, $E \geq 50$, changes in orbit class are infrequent since the size of the phase space region restricted by cantori is relatively small. However, for lower energies, these regions become larger and changes in orbit class more common.

This can be quantified by determining the minimum amplitude of friction and noise required to induce one or more changes in orbit class for a significant fraction of the orbits within time $t = 100$. Consider, e.g., the case $E = 20$ and $\Theta \sim E$. For $\eta$ much less than $10^{-9}$, the friction and noise are too weak to cause a significant number of orbits to change class. However, for $\eta \sim 10^{-9}$, friction and noise begin to become more important, and, already for $\eta$ as large as $10^{-6}$, as many as 50% of the noisy realisations for any initial condition can result in a change between filling and confined stochastic. Such changes are not accompanied by changes between regular and stochastic orbits, which, deterministically, are separated by $KAM$ tori, rather than cantori. Only for $\eta \geq 10^{-3}$ are any such changes observed.

What these simulations imply is that even very weak friction and noise, with characteristic time scale $t_R \geq 10^6$ characteristic crossing times $t_{cr}$ can have significant effects within a time as short as $\sim 10^2 \, t_{cr}$.

The authors acknowledge useful discussions with John Klauder and Ed Ott. SH was supported in part by the DOE and by AFOSR. HEK was supported by the NSF grant PHY92-03333. MEM was supported by the University of Florida. Computer time was made available through the Research Computing Initiative at the Northeast Regional Data Center (Florida) by the IBM Corp. and on the CM5, by the Advanced Computing Laboratory at Los Alamos National Laboratory.

# References

[1] R. S. MacKay, J. D. Meiss, and I. C. Percival, *Phys. Rev. Lett.* **52**, 697 (1984).

[2] Y.-T. Lau, J. M. Finn, and E. Ott, *Phys. Rev. Lett.* **66**, 978 (1991); J. D. Meiss and E. Ott, *Phys. Rev. Lett.* **55**, 271 (1985).

[3] H. E. Kandrup and M. E. Mahon, *Phys. Rev. E* **49**, 3735 (1994); *Astron. Astrophys.* (in press).

[4] P. Grassberger, R. Badii, and A. Politi, *J. Stat. Phys.* **51**, 135 (1988); M. A. Sepúlveda, R. Badii, and E. Pollak, *Phys. Rev. Lett.* **63**, 1226 (1989).

[5] M. A. Lieberman and A. J. Lichtenberg, *Phys. Rev. A* **5**, 1852 (1972).

[6] See, *e.g.*, M. Toda, R. Kubo, and N. Hashitsume, *Statistical Physics. II* (Springer-Verlag, Berlin, 1991).

[7] M. Toda, *J. Phys. Soc. Japan* **22**, 431 (1967); G. Contopoulos and C. Polymilis, *Physica* **24D**, 328 (1987).

[8] A. Greiner, W. Strittmatter, and J. Honerkamp, *J. Stat. Phys.* **51**, 95 (1988).

[9] D. Armbruster, J. Guckenheimer, and S. Kim, *Phys. Lett. A* **140**, 419 (1989).

[10] S. Habib, H. E. Kandrup, and M. E. Mahon, *Astrophys. J.* (to be submitted).

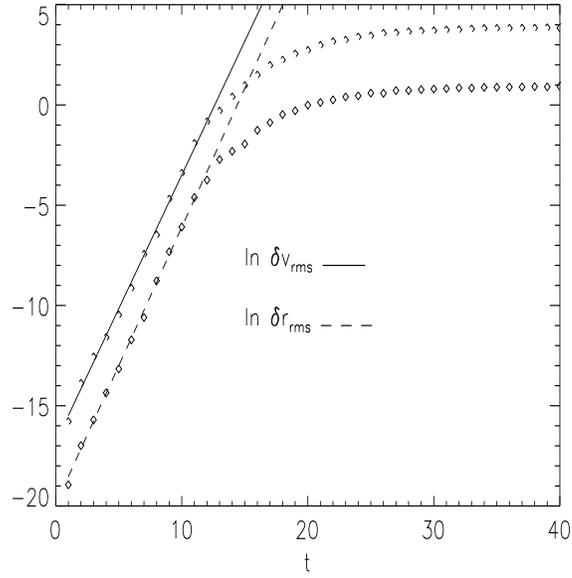

Figure 1: The early time exponential growth of $\delta r_{rms} \equiv (\delta x_{rms}^2 + \delta y_{rms}^2)^{1/2}$ (dashed curve) and the corresponding $\delta v_{rms}$ (solid curve) for $E = 30$ and $\eta = 10^{-9}$ is clearly apparent in this figure.

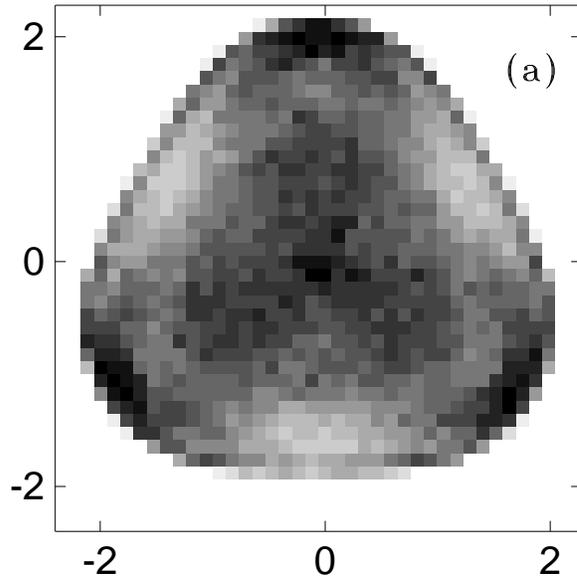

Figure 2: (a) A grey scale plot of the deterministic near-invariant $f_0(x,y)$ for $E = 30$, generated from a $40 \times 40$ binning. Darker shades represent higher densities.

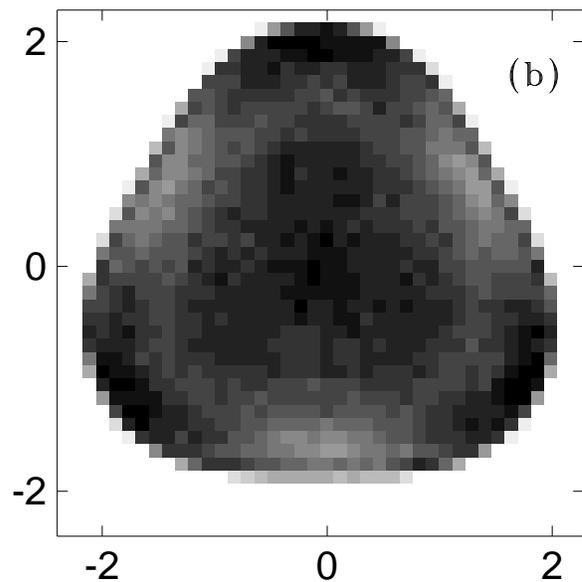

Figure 2: (b) The corresponding grey scale plot for the near-invariant $f_\eta(x,y)$ for $E = 30$ and $\eta = 10^{-6}$.

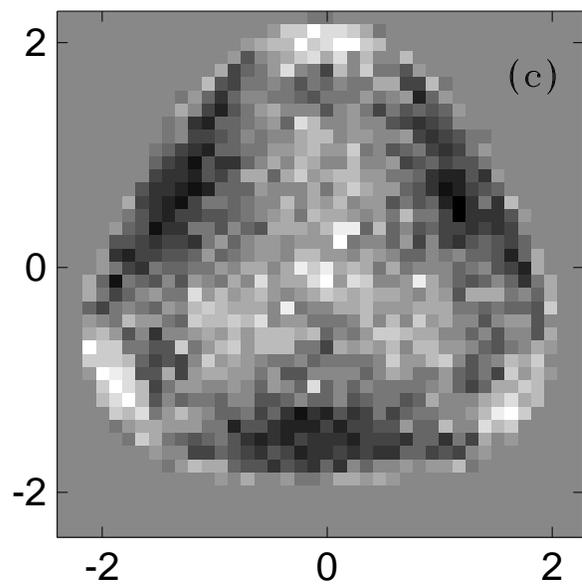

Figure 2: (c) The difference $f_0(x,y) - f_\eta(x,y)$.